\documentclass[twocolumn]{aastex63}

\usepackage{lineno}
\received{May 5, 2021}
\revised{June 2, 2021}
\accepted{August 6, 2021}

\submitjournal{ApJ}

\shorttitle{GAMA: The Merging Potential of BGGs}
\shortauthors{Banks et al.}

\graphicspath{{./}{figures/}}

\begin{document}

\title{Galaxy And Mass Assembly (GAMA) Survey: The Merging Potential of Brightest Group Galaxies}

\author{K. Banks}
\affiliation{School of Physics, University of New South Wales, NSW 2052, Australia}

\author{S. Brough}
\affiliation{School of Physics, University of New South Wales, NSW 2052, Australia}

\author{B.W. Holwerda}
\affiliation{Department of Physics and Astronomy, 102 Natural Science Building, University of Louisville, Louisville KY 40292, USA}

\author{A.M. Hopkins}
\affiliation{Australian Astronomical Optics, Macquarie University, 105 Delhi Rd, North Ryde, NSW 2113, Australia}

\author{{\'A}.R. {L{\'o}pez-S{\'a}nchez}}
\affiliation{Australian Astronomical Optics, Macquarie University, 105 Delhi Rd, North Ryde, NSW 2113, Australia}

\author{S. Phillipps}
\affiliation{HH Wills Physics Laboriatory, University of Bristol, Tyndall Avenue, Bristol, BS8 1TL, UK}

\author{K.A. Pimbblet}
\affiliation{School of Physics, Monash University, Clayton, Victoria 3800, Australia}
\affiliation{E.A.Milne Centre for Astrophysics, University of Hull, Cottingham Road, Kingston-upon-Hull, HU6 7RX, UK}

\author{A.S.G. Robotham}
\affiliation{ICRAR, The University of Western Australia, 35 Stirling Highway, Crawley, WA 6009, Australia}

\begin{abstract}

Using a volume-limited sample of 550 groups from the Galaxy And Mass Assembly (GAMA) Galaxy Group Catalogue spanning the halo mass range $12.8<\log[M_{h}/M_\odot]<14.2$, we investigate the merging potential of central Brightest Group Galaxies (BGGs). We use spectroscopically-confirmed close-companion galaxies as an indication of the potential stellar mass build-up of low-redshift BGGs, $z\leq0.2$. We identify 17 close-companion galaxies with projected separations $r_{p}<30~\rm kpc$, relative velocities $\Delta v\leq300~\rm km\,s^{-1}$, and stellar-mass ratios $M_{\text{BGG}}/M_{\text{CC}}\leq4$ relative to the BGG. These close-companion galaxies yield a total pair fraction of $0.03\pm0.01$. Overall, we find that BGGs in our sample have the potential to grow in stellar mass due to mergers by $2.2\pm1.5\%\,\rm Gyr^{-1}$. This is lower than the stellar mass growth predicted by current galaxy evolution models.

\end{abstract}

\keywords{galaxies: groups: general -- galaxies: evolution
-- galaxies: interactions}

\section{Introduction} \label{sec:intro}

Brightest Group Galaxies and Brightest Cluster Galaxies (BGGs, BCGs; e.g. \citealt{Collins09,Lidman12,Oliva14,Webb15}) are some of the most massive and most luminous galaxies observed in the Universe. They are identified within massive galaxy groups and clusters which are the largest gravitationally-bound structures in the Universe. BGGs and BCGs are often located at or near the centre of groups and clusters and are predicted to be the final stage of galaxy evolution. Due to their characteristically high luminosity, they can be observed at large distances in the Universe and are easily identified in cosmological simulations. These characteristics make them a particular focus of galaxy evolution studies.

BGGs and BCGs are predicted to increase in stellar mass in two phases: in the early stages of their evolution ($z\geq2$) star formation dominates their stellar mass growth, however, once star formation is quenched, their stellar mass growth is predicted to be dominated by galaxy mergers at $z\leq1$ (e.g. \citealt{DeLucia07,Laporte13,Contini14,Webb15,Gazaliasl16,Cerulo19,Cooke19}). It is evident from images that BCGs are often closely surrounded by other galaxies (e.g. \citealt{Schombert87}). These cluster members eventually spiral into the centre of the cluster potential due to dynamical friction, ultimately merging with the BCG located at the centre. Thus, the high mass of BCGs is attributed to their unique location near the centre of galaxy clusters (e.g. \citealt{Gunn72}). This mechanism is supported by observations (e.g. \citealt{ODea08,ODea10,Zhao17}).

There is a strong observed relationship between a BCG's stellar mass ($M_*$) and the mass of the dark matter halo it resides in ($M_{\text{halo}}$; i.e. its host cluster environment). More massive central galaxies are often associated with more massive haloes and the richness of their clusters/groups (i.e. the number density of galaxy members in a cluster/group; e.g. \citealt{Liu09,Zhao15}). The exact relationship has been examined in many studies (e.g. \citealt{Lin04,POpesso07,Brough08,Hansen09,Lidman12,Oliva14,Lavoie16,Kravtsov18}) which find the slope of $M_{*}$--$M^{b}_{\text{halo}}$ to be less than unity at $z<1$. This implies that, while the BCG and cluster grow together, the halo gains mass (by merging with other clusters and groups) significantly faster than the BCG.

The stellar mass growth of these galaxies can be measured observationally in two different ways. The first method involves directly examining stellar masses of samples at different redshifts (e.g. \citealt{Collins09,Lidman12,Oliva14,Liu15,Bellstedt16}). This is strongly dependent on the observed stellar mass-halo mass relationship. \cite{Lidman12} found that the stellar mass of BCGs increased by a factor of $1.8\pm0.3$ from $z=0.9$ to $z=0.2$. \cite{Oliva14} compared a sample of 883 galaxies divided into higher ($0.17\leq z\leq0.27$) and lower redshift bins ($0.09\leq z\leq0.17$) from the Galaxy and Mass Assembly Survey (GAMA; \citealt{Driver11}). They found no significant growth in the stellar mass of BGGs or BCGs over $\sim2\,$Gyr from $z=0.27$ to $z=0.09$. 

The second method predicts the stellar mass growth of these galaxies by examining the mass in close-companion galaxies available to merge with the central galaxy within a few Gyr (e.g. \citealt{McIntosh08,Liu09,Groenewald17}). 

Observational studies of the stellar mass growth of central galaxies due to mergers to-date have used samples that are either spectroscopically incomplete or use photometric redshifts (e.g. \citealt{McIntosh08,Liu09,Liu15,Groenewald17}). These studies will likely suffer from incompleteness or from line-of-sight contamination. Some studies (e.g. \citealt{Groenewald17}) use simulations to estimate a correction factor for the line-of-sight contamination. Others (e.g. \citealt{McIntosh08,Liu09,Liu15}) identify close-companion galaxies based not only on their projected separation from the central galaxy but also the presence of morphological disturbances in those galaxies. These corrections, however, can increase the uncertainties in these measurements.

The observations of the stellar mass build-up of central galaxies agree with simulations at high redshifts ($z>0.5$) but there is a large discrepancy between simulations and observations at lower redshifts (e.g. $z\leq0.5$; \citealt{Laporte13,Lidman13,Lin13,Contini14}). For example, both \cite{DeLucia07} and \cite{Laporte13} find that BGGs grow by a factor of $\sim1.8$ at $z<0.5$. This corresponds to an average fractional stellar mass growth of $\sim15\%$ per Gyr whereas observational studies such as \cite{McIntosh08,Liu09,Liu15,Oliva14} find significantly less growth.

In this paper we investigate the potential stellar mass growth of central BGGs between $0.07\leq z\leq0.20$ by analysing a volume-limited sample of groups selected from the Galaxy And Mass Assembly survey (GAMA; \citealt{Driver11}). GAMA offers a very large sample of galaxy groups that cover a wide range of total halo masses $(10^{10.3}$--$10^{15.0}$ M$_{\odot})$. GAMA's high spectroscopic completeness ($98.5\%$; \citealt{Liske15}) allows us to robustly examine the influence of merging close-companion galaxies on the stellar mass build-up of BGGs. 

In Section \ref{sec:GAMA} we outline the data source for this paper, the GAMA survey \citep{Driver11} and the different catalogues used within our analysis. Section \ref{sec:sample} details the selection of our volume-limited sample and describes the methods used to ensure a robust analysis. In Section \ref{sec:method} we describe the steps of our method and the calculations that allow us to investigate the stellar mass growth of BGGs in our sample. A discussion of our results and their comparison to other similar observational studies as well as semi-analytical models is presented in Section \ref{sec:discussion}. Finally, we summarise our conclusions in Section \ref{sec:conclusion}. Throughout this paper we assume a flat $\Lambda$CDM cosmology with $h=0.7$, $H_{0}=100h\,$km\,s$^{-1}$\,Mpc$^{-1}$, $\Omega_m=0.3$, and $\Omega_\Lambda=0.7$.

% ---------------NEW SECTION --------------- %
\section{The Galaxy And Mass Assembly (GAMA) Survey} \label{sec:GAMA}
The Galaxy And Mass Assembly (GAMA) survey is a wide-field, multi-wavelength galaxy redshift survey that probes the local Universe ($z\leq0.5$; \citealt{Driver11}). Spectroscopic observations were taken with the 3.9m Anglo-Australian Telescope in conjunction with the AAOmega multi-object spectrograph. The spectroscopic observations are highly complete with 98.5\% of galaxies having a robust distance measurement \citep{Hopkins13,Liske15}. GAMA also has a particular focus on high pair fraction completeness that is crucial for this analysis. This level of completeness was achieved by returning to observe each target area an average of 10 times \citep{Robotham10}.

The GAMA survey consists of $\sim$300,000 galaxies with a magnitude limit of $r<19.8$ mag over $\sim286\text{ deg}^{2}$ across five regions of the sky. This study analyses the galaxies within the three main survey regions, G09, G12, and G15, known as the equatorial regions \citep{Driver11}. Survey data is made available in Data Management Units (DMUs) that collate measurements from different analyses. The group-finding and stellar mass DMUs are of particular use to this analysis and are described in the following sections.

% ---------------NEW SUBSECTION --------------- %
\subsection{The GAMA Galaxy Group Catalogue (G$^3$C)} \label{G3C}
The GAMA Galaxy Group Catalogue (G$^3$C) is the catalogue of all of the galaxy groups defined in the GAMA survey \citep{Robotham11}. We use version 10 of the catalogue here and briefly describe the key measurements. 

The G$^{3}$C uses a friends-of-friends (FoF) algorithm to identify groups. The halo mass is estimated using the measured values of the halo velocity dispersion ($\sigma$) and the group radius ($R$). The halo velocity dispersion is calculated using the \textit{gapper} estimator introduced by \cite{Beers90} which is developed to be robust to outliers in smaller number samples. The group radius used is that which contains 50\% of galaxies in the group (Rad$_{50}$). This is chosen such that the group radius is robust against potential interloping galaxies \citep{Robotham11}. For a stable system, the halo mass equates to $M_{\text{halo}}=A\sigma^{2}R$. In GAMA, the constant $A$ is calculated by comparison between the group-finder outputs from observations and mock observations of simulations \citep{Robotham11}, and the total halo mass of a group is estimated as $M=10.0\sigma^{2}R$. We also estimated the impact of determining halo masses of our groups using the weak-lensing-calibrated halo mass scaling relations in \cite{Viola15} and find that these do not qualitatively change our conclusions. 

Three approaches were considered to determine the central galaxy of GAMA groups in \cite{Robotham11}. The first approach determined the centre of the cluster to be the centre of light which is a good proxy for centre of mass. The second approach was an iterative process where the centre of light was derived at each step from the $r_{AB}$-band luminosity of all the galaxies identified within the group. The most distant galaxy from the centre of light was rejected. This process was repeated until two galaxies remain, of which the brighter $r_{AB}$-band galaxy is used as the group centre. The third approach simply defined the group centre by the brightest member of the group, i.e. the BGG. \cite{Robotham11} found that the iterative method produced the most robust estimate of the central galaxy of the group and we therefore use that here.

% ---------------NEW SUBSECTION --------------- %
\subsection{GAMA Stellar Mass Catalogue} \label{GAMA_stelM}
The stellar masses of galaxies in the GAMA survey are estimated by \cite{Taylor11} by fitting model spectral energy distributions to Sloan Digital Sky Survey (SDSS; \citealt{York00}) \textit{ugriz} imaging reprocessed by the GAMA team \citep{Hill11}. When observing galaxies at varying distances with an aperture of a fixed size, different fractions of a galaxy's luminosity will be observed. The flux of each galaxy is measured within a flexible circular aperture where the size of the aperture is determined by the observed radial surface brightness profile of the galaxy. Since only a fraction of each galaxy's luminosity is considered for the stellar mass estimate, a linear scale factor, \texttt{fluxscale}, is calculated to account for the unobserved luminosity of each galaxy from the ratio between the aperture $r$-band flux and the total $r$-band flux inferred from fitting a single Sersic profile truncated at $10\,R_e$ \citep{Kelvin12}. We apply this scale factor in this work.

% ---------------NEW SECTION --------------- %
\section{Sample Selection} \label{sec:sample}
The G$^3$C equatorial regions contain $23,654$ groups with $75,029$ constituent members. Estimated parameters for groups within the G$^3$C, such as the projected radius Rad$_{50}$, velocity dispersion $\sigma$, and halo mass $M_{\text{halo}}$, are robust for groups that contain five or more galaxy members ($N_{\text{FoF}}$; \citealt{Robotham11}). Hence, we only include groups with five or more galaxy members in our analysis, $N_{\text{FoF}}\geq5$. This reduces our sample to $2,754$ groups. Furthermore, five groups with $N_{\text{FoF}}\geq5$ have an estimated velocity dispersion $\sigma=0$ since the error in the raw velocity distribution $\sigma_{\text{err}}$ is larger than the raw velocity distribution $\sigma_{\text{gap}}$. Hence, we do not include these five groups in this analysis.

In this study, we wish to examine groups where the identified BGG is at the centre of the group, therefore, we choose groups in our sample where the BGG is also identified as the iterative central galaxy. Selecting groups where the BGG is also identified as the iterative central gives us a sample of $2,363$ groups.

There are two BGGs within the G$^3$C that do not have a measured stellar mass within the Stellar Mass catalogue, likely due to contamination in one of the images. We exclude the groups that host these two BGGs from our sample. This leaves us with a sample of $2,361$ groups. 

% ---------------NEW SUBSECTION --------------- %
\subsection{BGG Selection} \label{Volume}
The GAMA survey is apparent-magnitude limited, we therefore select a volume-limited sample to ensure a robust analysis. 

    \begin{figure}[t!]
        \centering
        \includegraphics[width=0.48\textwidth]{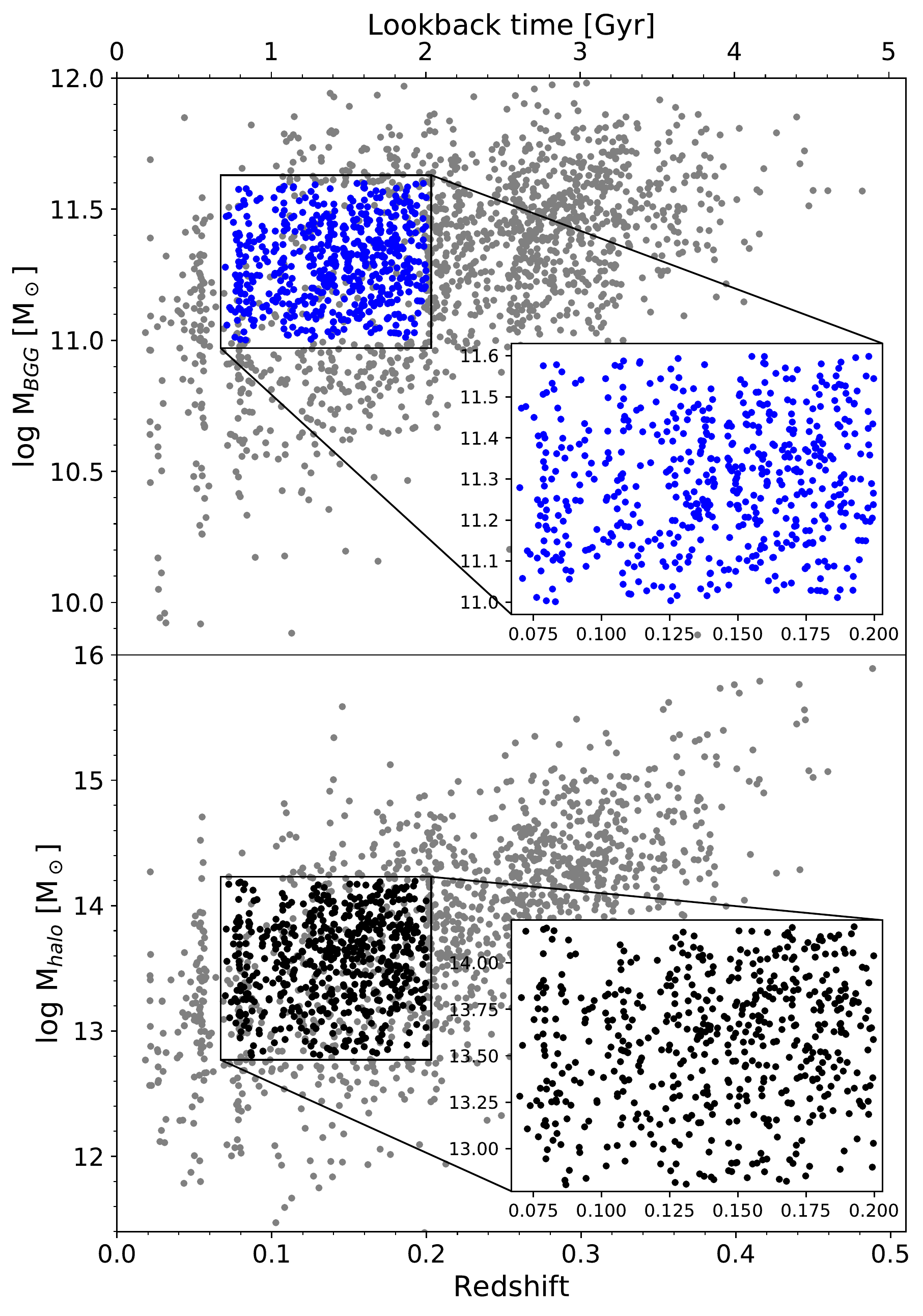}
        \caption{Selection of BGG sample as a function of stellar mass and redshift (top panel). Selection of BGG sample as a function of halo mass and redshift (bottom panel). The grey points represent the preliminary G$^3$C sample without BCG/halo mass selection, the blue points in the top panel and the black points in the bottom panel illustrate the final group sample selection with respect to BCG stellar mass and halo mass respectively.}
        \label{fig:BGG_Halo_sample}
    \end{figure}

The distribution of BGG stellar mass with redshift is shown in the top panel of Fig. \ref{fig:BGG_Halo_sample}. The density of BGGs with $z\sim0.25$ is lower compared to the rest of the sample, due to the 5577 Å sky line passing through key spectral features. We therefore limit our sample to groups with redshifts $z<0.2$. We also impose a lower limit on the redshift, $z\geq0.07$, due to the low volume sampled below this.

We define a minimum BGG stellar mass limit across our sample such that this minimum stellar mass is observed for galaxies of all colours across our redshift range. At $z=0.2$ this minimum complete BGG stellar mass is $10^{11.0}\,$M$_\odot$ which we apply to our sample. We also apply an upper limit, $10^{11.6}$\,M$_\odot$, on the stellar mass of BGGs to remove unrealistically massive BGGs, which likely occur as a result of image contamination. This volume limit yields a sample of 640 groups. The distribution of halo mass is also affected by GAMA's apparent-magnitude selection limit. We apply a volume limit to this too, limiting the halo mass of groups to a lower limit of $M_h=10^{12.8}$\,M$_\odot$ and an upper limit of $10^{14.2}$\,M$_\odot$. This selection results in a volume-limited sample of 550 groups and is illustrated in Fig. \ref{fig:BGG_Halo_sample} with BGG stellar mass in the top panel and group halo mass in the bottom panel.

    \begin{figure}[t!]
        \centering
        \includegraphics[width=0.48\textwidth]{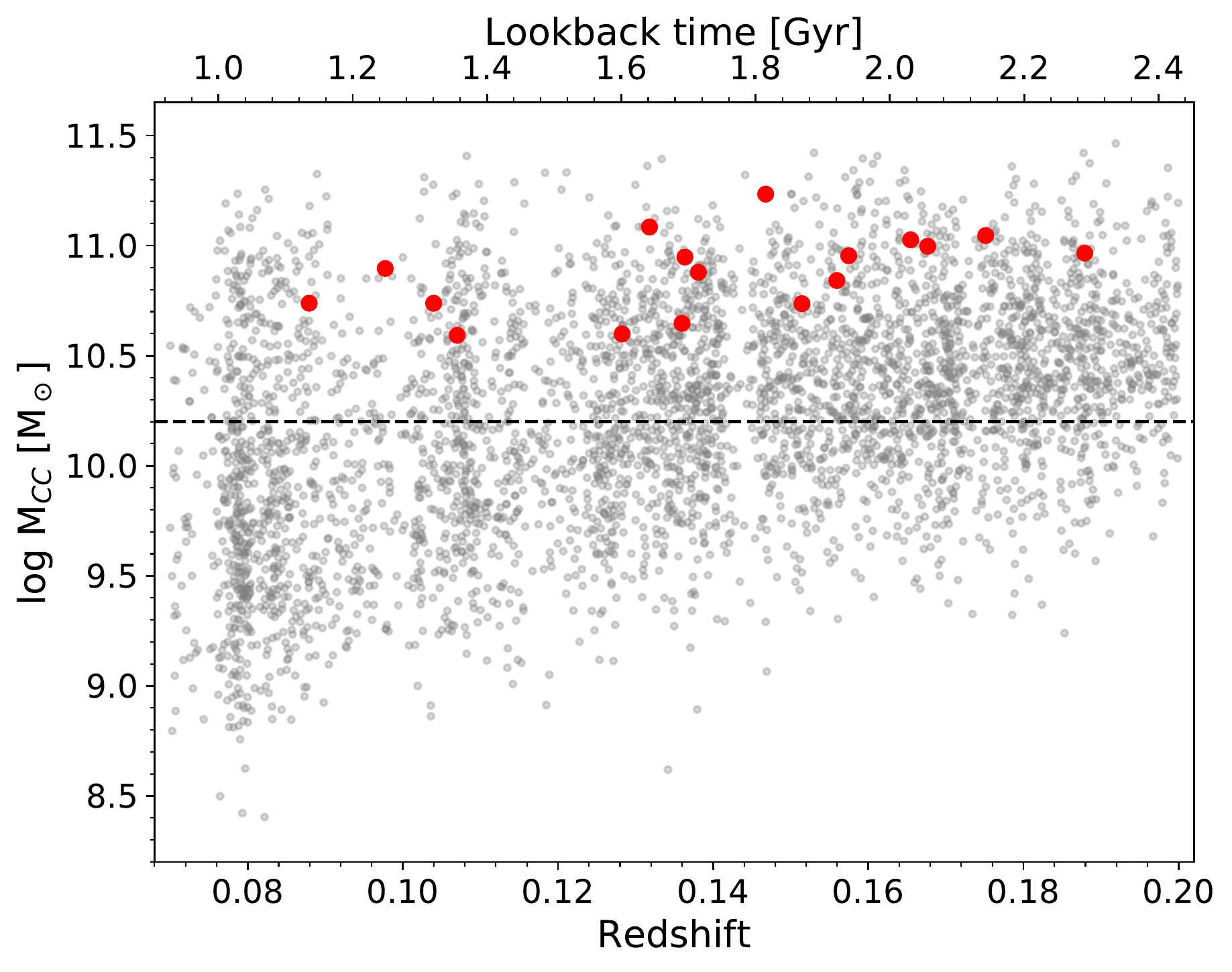}
        \caption{Selection of close-companion galaxies illustrated with respect to redshift and companion galaxy stellar mass. The grey points represent the non-central galaxy members of groups in our sample and the red points illustrate the close-companion galaxies identified from our selection criteria. The dotted black line at log $M_{CC}=10.2$\,M$_\odot$ illustrates the lower stellar mass limit of our sample of close-companion galaxies.}
        \label{fig:CC_sample}
    \end{figure}
    
    \begin{figure}[t!]
        \centering
        \includegraphics[width=0.48\textwidth]{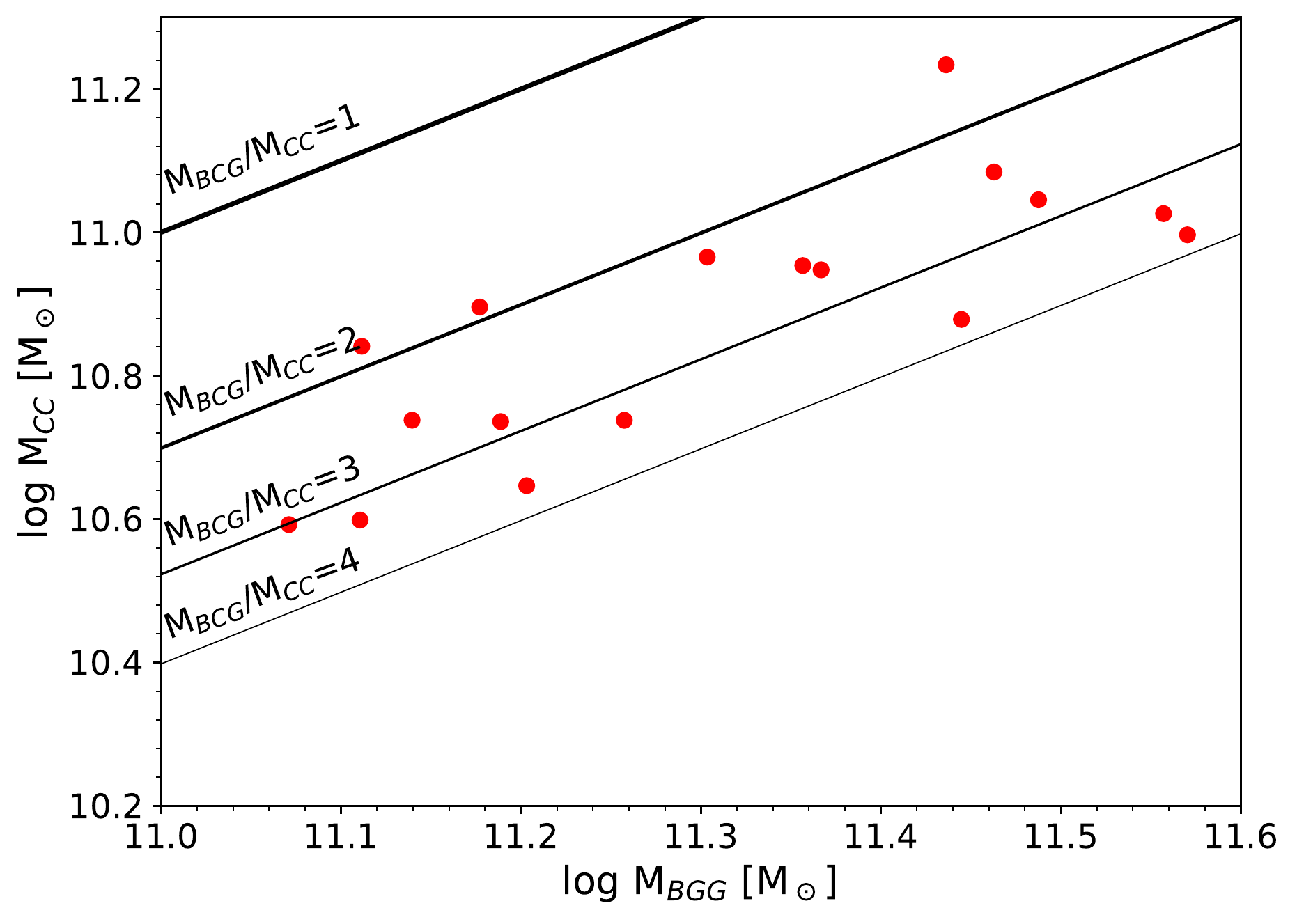}
        \caption{Distribution of close-companion galaxies within our sample with respect to BGG stellar mass. Each black line represents a BGG-to-CC mass ratio equipotential, with the thickest line representing a mass ratio equivalent to 1 and the weakest line equivalent to 4. This illustrates that the companions in our sample are spread uniformly across stellar mass ratios.}
        \label{fig:CC_ratio}
    \end{figure}

% ---------------NEW SUBSECTION --------------- %
\subsection{Close-Companion Galaxy Selection} \label{sec:CCs}
Our sample of 550 groups contains 4,207 non-central member galaxies that have the potential to merge with their BGG (illustrated by the grey points in Fig. \ref{fig:CC_sample}). The distribution of these galaxies in redshift is similarly affected by GAMA's apparent magnitude selection limit. We place a conservative initial volume limit of $M^{*}=10^{10.2}\,\text{M}_\odot$ allowing for the range of galaxy colours at GAMA's $m_r=19.8\,$ mag limit at our $z\sim0.2$ limit. This yields a maximum possible BGG-to-companion stellar mass ratio $M_{\text{BGG}}/M_{\text{CC}}\sim6$ when compared to the least massive BGGs in our sample (i.e. $\sim10^{11.0}\,$M$_\odot$).

Within the literature (e.g. \citealt{DeLucia07,McIntosh08,Liu09,Robotham14,Groenewald17}) a close-companion galaxy with a BGG-to-companion stellar mass ratio $M_{\text{BGG}}/M_{\text{CC}}\leq4$ is considered a major merger, whereas those with stellar mass ratios $>4$ are considered minor mergers. The lower stellar mass limit imposed on our sample of potential close-companion galaxies does not produce a robust sample of minor merger candidates. Henceforth, we focus only on major mergers with stellar mass ratios $M_{\text{BGG}}/M_{\text{CC}}\leq4$.

A galaxy may be considered a close companion to a BGG if it has a small projected separation ($r_p$) and relative velocity ($\Delta v$) to the BGG. Taking into account both of these parameters ensures an unbiased sample with less of the line-of-sight contamination that may be present in purely photometric samples \citep{Kitzbichler08}.

A recent analysis of the \textsc{Illustris-1} simulation \citep{Vogelsberger14,Ventou19} suggests that major close-companion galaxies at low redshift ($z<0.5$) with a projected distance $r_{p}\leq25\,\text{kpc}$ and a velocity of $\Delta v\leq150\,\text{km}\,\text{s}^{-1}$ relative to the BGG with stellar mass $>10^{9.5}\,$M$_\odot$ have a 75\% chance of merging by $z=0$. \cite{Ventou19} suggest the following selection criteria for selecting spectroscopic close-companion galaxies that are likely to merge: $r_p\leq50\,$kpc and $\Delta v\leq300\,$km\,s$^{-1}$. However, numerous observational studies (e.g. \citealt{McIntosh08,Liu09,Robotham14,Groenewald17}) apply a more conservative projected separation of $r_p\leq30$\,kpc.

To construct a robust sample of close-companion galaxies we follow the recommendations of \cite{Ventou19} and identify close-companion galaxies as galaxies with $\Delta v\leq300$\,km\,s$^{-1}$ and use the conservative projected separation limit $r_p\leq30$\,kpc. This selection criteria results in 17 close-companion galaxies. We note here that when we expand the sample to $r_p\leq50\,$kpc, the sample increases to 31 close-companion galaxies. To be consistent with the literature we only consider those within $r_p\leq30\,$kpc for the remainder of our analysis. These close-companion galaxies are represented by the red points in Fig. \ref{fig:CC_sample}.

Figure \ref{fig:CC_ratio} illustrates the distribution of companion stellar mass with respect to BGG stellar mass and demonstrates equivalent BGG-to-CC stellar mass ratios. Each black line represents stellar mass ratios with the thickest line representing $M_{\text{BGG}}/M_{\text{CC}}=1$ and the weakest line equivalent to 4. This illustrates that the companions within our sample are distributed uniformly between stellar mass ratios $1\leq M_{\text{BGG}}/M_{\text{CC}}<4$.

% ---------------NEW SECTION --------------- %
\section{Stellar Mass Growth of BGGs} \label{sec:method}
We are analysing the potential for stellar mass growth of BGGs over the redshift range $0.07\leq z\leq0.20$. We also investigate whether there is any dependence on the halo mass of their host cluster which we use as a tracer of the environment of the BGGs.

\subsection{Pair Fraction}
We investigate the fraction of close-companion galaxies within our sample. This is also known as the pair fraction and is defined as:
\begin{equation}
    f_{\text{pair}}=\frac{N_{\text{CCs}}}{N_{\text{BGGs}}}
\end{equation}
where $N_{BGGs}$ is the total number of BGGs in the sample and $N_{CCs}$ is the total number of identified close-companion galaxies. The $1\sigma$ uncertainties used throughout are calculated using the binomial confidence interval described in \cite{Cameron11} because it is robust to the small sample sizes present in our analysis. 

Given that we have a sample of 550 BGGs and identified 17 close-companion galaxies within our sample of non-central galaxies, this yields a pair fraction of $f_{\text{pair}}=0.03\pm0.01$. However, as mentioned in Section \ref{sec:CCs}, simulations show that close-companion galaxies with $r_p\leq25\,$kpc and $\Delta v\leq150\,$km\,s have a 75\% chance of merging by $z=0$ so the pair fraction calculated here is an upper limit. The pair fraction is henceforth used to calculate the maximum potential stellar mass growth of the BGGs in our sample.

\begin{figure}
        \centering
        \includegraphics[width=0.48\textwidth]{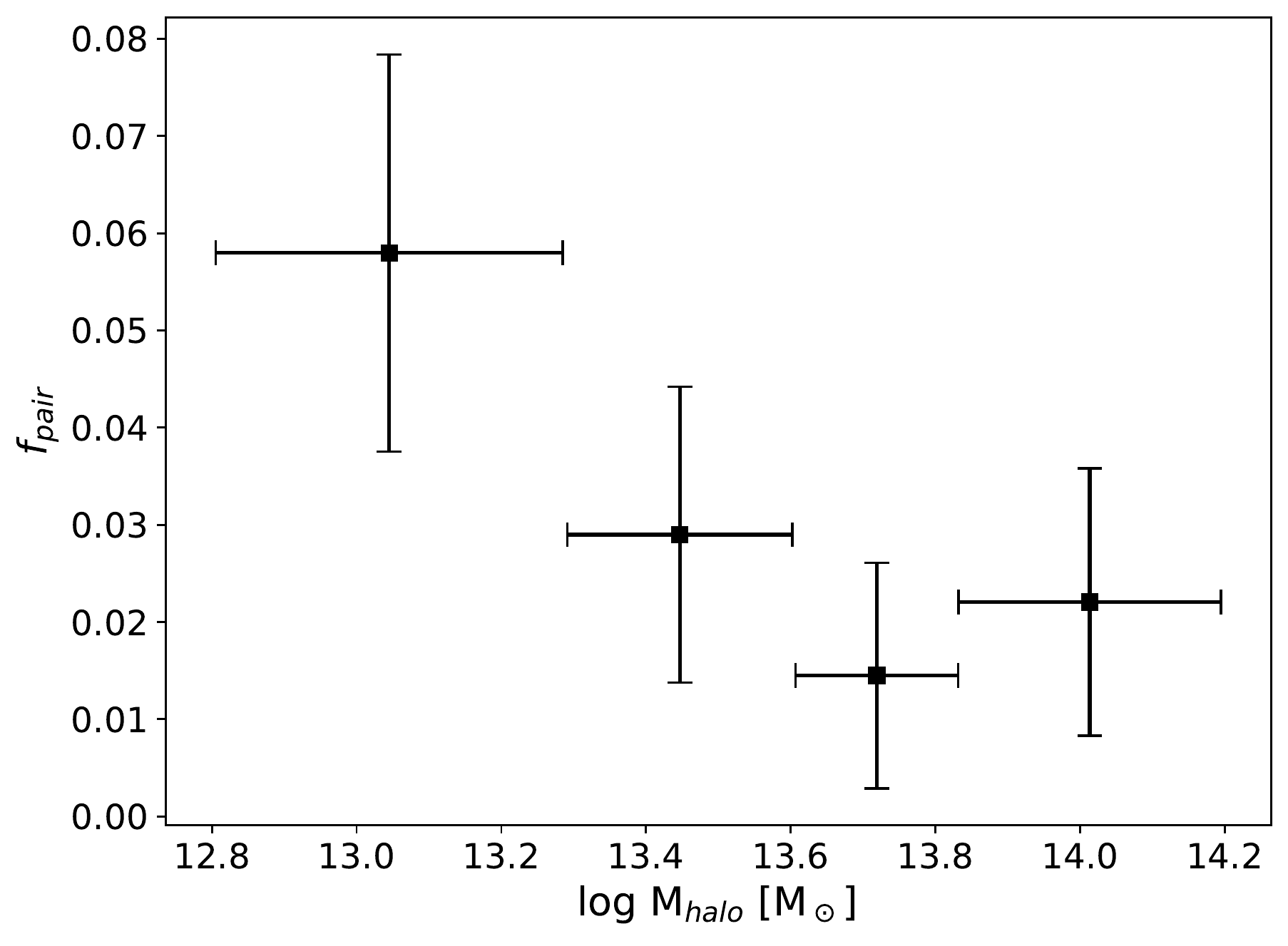}
        \caption{BGG pair fraction as a function of halo mass. We find no statistically significant dependence of the BGG pair fraction on halo mass over our halo mass range $10^{12.8}\,\text{M}_{\odot}\leq M_{\text{halo}}\leq10^{14.2}\,\text{M}_\odot$. We do note, however, a systematic decrease of the BGG pair fraction with increasing halo mass. This is also present with a less conservative limit of $r_{p}\leq50\,$kpc on the close-companion galaxies.}
        \label{fig:halo_pair}
    \end{figure}

We also investigate the influence of halo mass on the BGG pair fraction and illustrate this in Fig. \ref{fig:halo_pair}. While there is no statistically significant dependence in this halo mass range (i.e. $10^{12.8}\,\text{M}_{\odot}\leq M_{\text{halo}}\leq10^{14.2}\,\text{M}_\odot$), we do note a systematic decrease of the BGG pair fraction with increasing halo mass. We also investigated this relationship with the less conservative limit of $r_{p}\leq50\,$kpc and found a similar result.

% ---------------NEW SUBSECTION --------------- %
\subsection{Merger Rate}
The next step in determining the potential stellar mass growth of BGGs is to calculate their merger rate, which requires knowledge of the time it takes for a close-companion galaxy galaxy to merge with the BGG, i.e. the merging timescale. A commonly used merging timescale in galaxy merging literature is that derived by \cite{Kitzbichler08} using the Millenium Simulation \citep{Springel05}.

\cite{Kitzbichler08} find that at redshifts $z\ll1$, the average merging timescale derived for a close-companion galaxy galaxy with stellar mass $>5\times10^{9}h^{-1}\,\text{M}_\odot$, within $\Delta v\leq300\,$km\,s$^{-1}$ and a projected radius $r_{p}\leq50\,\rm kpc$ is given by:
\begin{equation}
    \langle T_{\text{merge}}\rangle=T_0\frac{r_p}{r_0}
    \left(\frac{M_*}{M_0}\right)^{-0.3}
    \left(1+\frac{z}{8}\right)
    \label{eq:Kitz}
\end{equation}
where $T_0=2.2\,$Gyr, $r_0=50h^{-1}\,$kpc, $M_0=4\times10^{10}h^{-1}\,$M$_\odot$, and $z$ is the median redshift of the cluster. Using this method the close-companion galaxies identified in this work are predicted to merge with their respective BGGs in $0.53\pm0.19\,$Gyr on average. 

Other recent investigations into the average merger rate of galaxies use an observability timescale (e.g. \citealt{Lotz08,Lotz11,Mundy17,Duncan19}). This observability timescale is the average timescale during which merging galaxies can be observed depending on the method used to identify the merger (e.g. close galaxy pair selection utilised in this work). \cite{Lotz11} derive this quantity for a number of merger selections. They suggest that the timescale for major close-companion galaxies identified through pair selection is $\sim0.33\,$Gyr for $5h^{-1}<r_p<20h^{-1}\,$kpc, and $\sim0.63\,$Gyr for $10h^{-1}<r_p<30h^{-1}\,$kpc, which is consistent with the merging timescale calculated for our close-companion galaxies using equation (\ref{eq:Kitz}). In order to provide a robust comparison to previous analyses (e.g. \citealt{McIntosh08,Liu09,Groenewald17}) of the stellar mass growth of BGGs we use the timescales predicted by \cite{Kitzbichler08} which are also utilised in these previous studies.

The average merger rate of a sample of BGGs, i.e., the number of mergers per BGG per Gyr is then calculated as follows:
\begin{equation}
    \langle R_{\text{merge}}\rangle=\frac{f_{\text{pair}}}{\langle T_{\text{merge}}\rangle}
\end{equation}
The uncertainty on $\langle R_{\text{merge}}\rangle$ is calculated using the standard propagation of uncertainties.

The close-companion galaxies in our sample of groups merge with their respective BGGs in $0.53\pm0.19$\,Gyr on average. Therefore, the BGGs in our sample experience on average $0.06\pm0.03$ mergers per Gyr.

% ---------------NEW SUBSECTION --------------- %
\subsection{Average Stellar Mass Growth}
The potential stellar mass growth rate of BGGs, $\Delta M_*$ (M$_\odot$\,Gyr$^{-1}$) is calculated using a modified version of equation (7) from \cite{Groenewald17}. Their equation determines the overall growth of BGGs evolved from a particular redshift to $z=0$. In this study we are interested in the potential average mass growth of BGGs at low redshifts, so we modify their equation such that the result is the average stellar mass growth per Gyr. Our modified equation is as follows:
\begin{equation}
    \Delta M_*=\langle R_{\text{merge}}\rangle\times\langle M_*\rangle_{\text{CC}}
\end{equation}
where $\langle M_*\rangle_{\text{CC}}$ is the average stellar mass of the close-companion galaxies in our sample in M$_{\odot}$.

The close-companion galaxies in our sample have an average stellar mass of $\langle M_*\rangle_{\text{CC}}=8.18\pm3.34\times10^{10}\,\text{M}_\odot$, which contribute a total stellar mass of $\Delta M_{*}=0.47\pm0.30\times10^{10}\,\text{M}_{\odot}$ per Gyr.

% ---------------NEW SUBSECTION --------------- %
\subsection{Fractional Stellar Mass Growth}
We calculate the fractional contribution of mergers toward the stellar mass growth of BGGs per Gyr. We modify Equation (8) from \cite{Groenewald17} to calculate the fractional contribution made by merging close-companion galaxies in our sample but not incorporating their growth to $z=0$. This is defined per Gyr, by:
\begin{equation}
    F=\frac{\Delta M_*}{\langle M_*\rangle_{\text{BGG}}(t=0)+\Delta M_*}
\end{equation}
where $\langle M_*\rangle_{\text{BGG}}(t=0)$ is the average stellar mass of BGGs in the sample at the present day. While BGGs do grow in stellar mass from the formation of stars it is rare to find star-forming BGGs at $z\leq0.5$ ($<1\%$; e.g. \citealt{Liu12,FraserMcK14,Webb15,Groenewald17,Cerulo19}). Those BGGs that are forming stars grow by $\sim1-3\%$ in stellar mass from star formation (e.g. \citealt{Liu12}). Furthermore, it is estimated that the contribution of stellar mass to BGGs via minor mergers is just as significant as that through major mergers (e.g. \citealt{Edwards12}), however, we cannot robustly measure the stellar mass growth of BGGs due to minor mergers and so we focus here on their growth only via major mergers.

We find that BGGs in our sample spanning the redshift range $0.07\leq z\leq0.20$ grow in stellar mass due to major mergers by $2.19\%\pm1.52\%\,\text{Gyr}^{-1}$, assuming that all of the stellar mass of a merging close-companion galaxy is accreted onto the BGG. This is similar to the predicted stellar mass growth from star formation of star forming BGG ($\sim1-3\%$ e.g. \citealt{Liu12})

    \begin{table*}[t]
        \centering
        \begin{tabular}{c c c c c c c c}
            Redshift & $\langle M_{\text{BGG}} \rangle$ & $\langle M_{\text{CC}}\rangle$ & $f_{\text{pair}}$ & $\langle T_{\text{merge}}\rangle$ & $\langle R_{\text{merge}}\rangle$ & $\Delta M_*$ & $F$ \\
             & [$10^{10}\,\text{M}_\odot$] & [$10^{10}\,\text{M}_\odot$] & & [Gyr] & [Gyr$^{-1}$] & [$10^{10}\,\text{M}_\odot\,$Gyr$^{-1}$] & [\%\,Gyr$^{-1}$] \\
            \hline
            $0.07\leq z\leq 0.2$ & $21.17\pm7.48$ & $8.18\pm3.34$ & $0.03\pm0.01$ & $0.53\pm0.19$ & $0.06\pm0.03$ & $0.47\pm0.30$ & $2.19\pm1.52$
            \\
        \end{tabular}
        \caption{Results. Column 1 is the redshift range of our sample. Columns 2 and 3 indicate the average stellar mass of the BGGs and close-companion galaxy (CCs) galaxies in our sample respectively. Columns 4, 5, and 6 illustrate the pair fraction, average merging timescale, and average merger rate for the BGGs in our sample respectively. Finally, columns 7 and 8 are the stellar mass growth of BGGs in our sample in solar masses per Gyr and fractional stellar mass growth per Gyr respectively.}
        \label{tab:results}
    \end{table*}

% ---------------NEW SECTION --------------- %
\section{Discussion} \label{sec:discussion}
We have presented here an analysis of the potential stellar mass growth of BGGs in the GAMA survey from close-companion galaxies. In this section we compare our results to the literature. 

\subsection{BGG Pair Fraction} \label{BGG_fpair}
\begin{figure}[t!]
    \centering
    \includegraphics[width=0.48\textwidth]{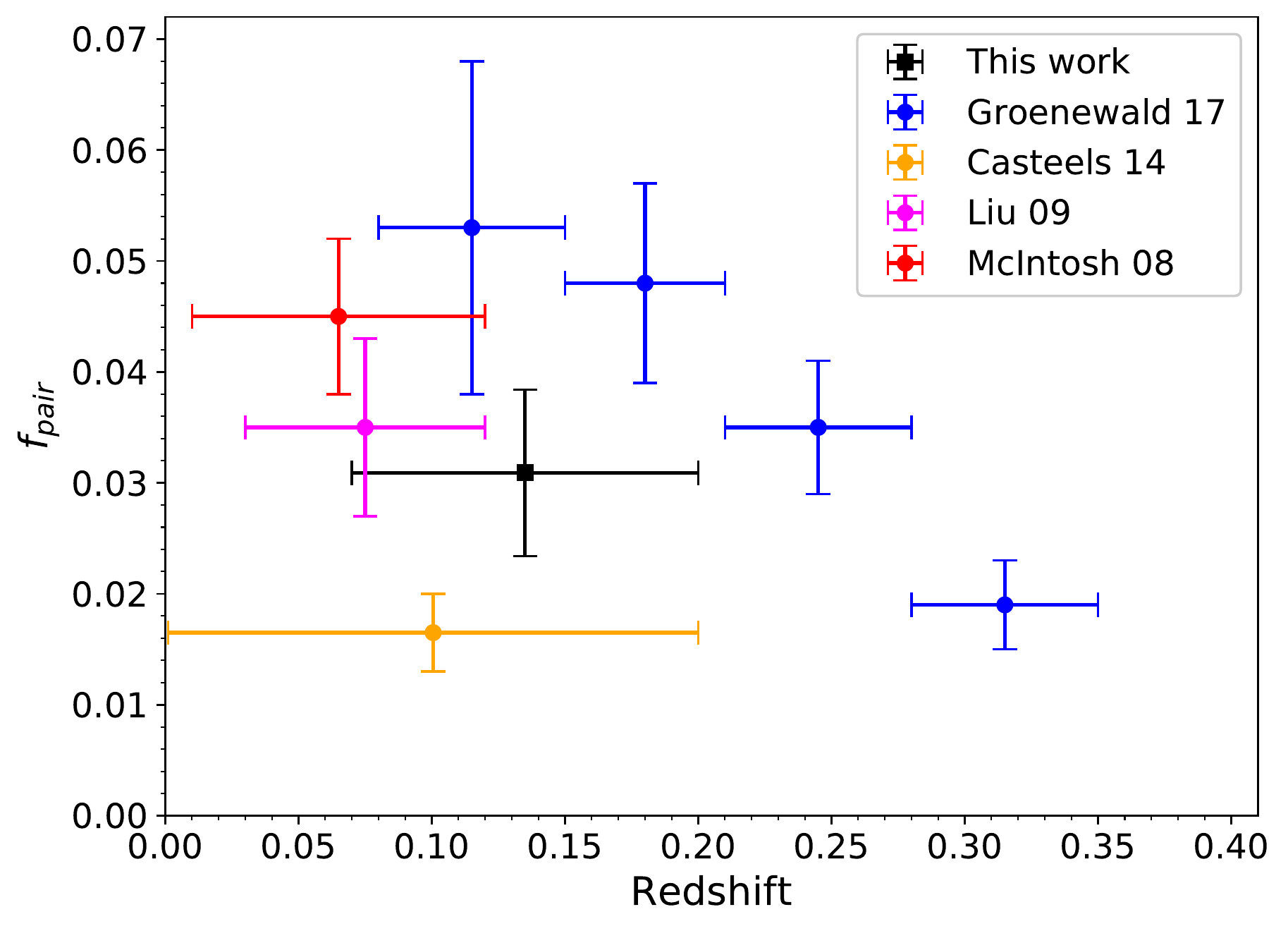}
    \caption{Major merger pair fraction considering close-companion galaxies within 30\,kpc and with mass ratios $\leq4$. Our result is presented as the black point. Other published pair fractions are plotted at the mean redshift of each sample while the horizontal error bars indicate the redshift range of the sample. The BGG pair fraction we calculate in our sample agrees well with previous results.}
    \label{fig:fpair_compare}
\end{figure}

In Fig. \ref{fig:fpair_compare} we compare our pair fraction to earlier studies. Both \cite{McIntosh08} and \cite{Liu09} have studied the BGG pair fraction at $z\leq0.12$. \cite{McIntosh08} investigated the incidence of major mergers with mass ratios $\leq4$ in the Sloan Digital Sky Survey (SDSS). They used a volume-limited sample of 845 groups with halo masses $>2.5\times10^{13}$ M$_\odot$ and refined their search for merger candidates within 30\,kpc by visually inspecting their sample of 221 galaxy pairs for the presence of morphological features associated with merging events. This was done in lieu of setting a limit on the relative velocities of the close-companion galaxies due to the purely photometric sample used in their analyses. From this they found a pair fraction of $f_{\text{pair}}=0.045\pm0.007$. This is represented by the red point in Fig. \ref{fig:fpair_compare}.

\cite{Liu09} similarly searched for ongoing major mergers ($\leq4$) in a sample of BGGs from the SDSS C4 cluster catalogue \citep{Miller05} with redshifts $0.03\leq z\leq0.12$. They also searched for close-companion galaxies within 30\,kpc that showed significant signs of interaction in the form of significant asymmetry in residual images. They concluded that 18 of their 515 BGGs were involved in major mergers, i.e. $f_{\text{pair}}=0.035\pm0.008$ indicated in Fig. \ref{fig:fpair_compare} by the magenta point.

The blue points in Fig. \ref{fig:fpair_compare} represent the pair fraction calculated in \cite{Groenewald17}. They studied more massive groups ($M_h>2.2\times10^{15}$\,M$_\odot$) constructed from the redMaPPer catalogue \citep{Rykoff14}. They split their sample into four redshift ranges, the first two of which overlap with the redshift covered by our sample. These two low redshift ranges result in a pair fraction $f_{\text{pair}}\sim0.05$.

We calculate a major merger pair fraction of $f_{\text{pair}}=0.03\pm0.01$ over similar redshifts to those in \cite{McIntosh08,Liu09} and \cite{Groenewald17}. This is in agreement with these earlier studies.

 \citealt{Casteels14} also investigated the major merger pair fraction in the GAMA survey at redshifts $0.001<z<0.2$. They examined the mass dependent major merger rate of GAMA galaxies with stellar masses $10^{8.0}<M_*<10^{11.5}\,$M$_{\odot}$ and found the major merger pair fraction to be consistent at $\sim0.013-0.02$ between $10^{9.5}<M_*<10^{11.5}\,$M$_{\odot}$ an major merger pair fraction is approximately half the major merger pair fraction calculated in this work (i.e. $0.03\pm0.01$), however, we note that their sample is not limited to central galaxies and so we do not directly compare it to our result in Fig. \ref{fig:fpair_compare}.

We also investigate the role the BGG's environment plays on the pair fraction (see Fig. \ref{fig:halo_pair}). \cite{Liu09} also examined the relation between the fraction of BGGs involved in major mergers and the richness of the cluster, defined as the number of cluster members within a 1$h^{-1}$\,Mpc radius centred on the BGG. The pair fraction of BGGs in \cite{Liu09} appeared to increase with increasing cluster richness with a BGG pair fraction of $\sim0.01$ at a richness of $\sim15$ up to $\sim0.055$ at a richness of $\sim45$. While our total pair fractions across halo mass are in agreement with those in \cite{Liu09}, Fig. \ref{fig:halo_pair} shows that we find that the major merger pair fraction tends to decrease with halo mass, however, we note that relationship is not statistically significant.

% ---------------NEW SUBSECTION --------------- %
\subsection{Merger Rate} \label{Rmerge}
The BGG pair fraction and the mean merging timescale of close-companion galaxies are key ingredients in the calculation of the BGG merger rate. We have estimated the merging timescale of close-companion galaxies in our sample using the merging timescale derived in \cite{Kitzbichler08}. The merging timescale is largely influenced by the stellar mass of the in-falling close-companion galaxies and their projected separation from the BGG. Our sample of close-companion galaxies range in stellar mass between $10.5<\log[M_*/$M$_\odot]<11.2$ and have projected separations within 30\,kpc. The mean merging timescale for all of the close-companion galaxies in our sample is $0.53\pm0.19$\,Gyr.

A pair fraction of $0.03\pm0.01$ and a mean merging timescale of $0.53\pm0.19$\,Gyr yields an average merger rate of $0.06\pm0.03$\,Gyr$^{-1}$. This implies that the BGGs in our sample experience on average one major merger every 16\,Gyr since $z=0.2$, significantly longer than the age of the Universe. This is consistent with simulations (e.g. \citealt{Hopkins10}) and observations at similar redshifts (e.g. \citealt{McIntosh08,Edwards12,Liu12}).

\begin{figure}
    \centering
    \includegraphics[width=0.48\textwidth]{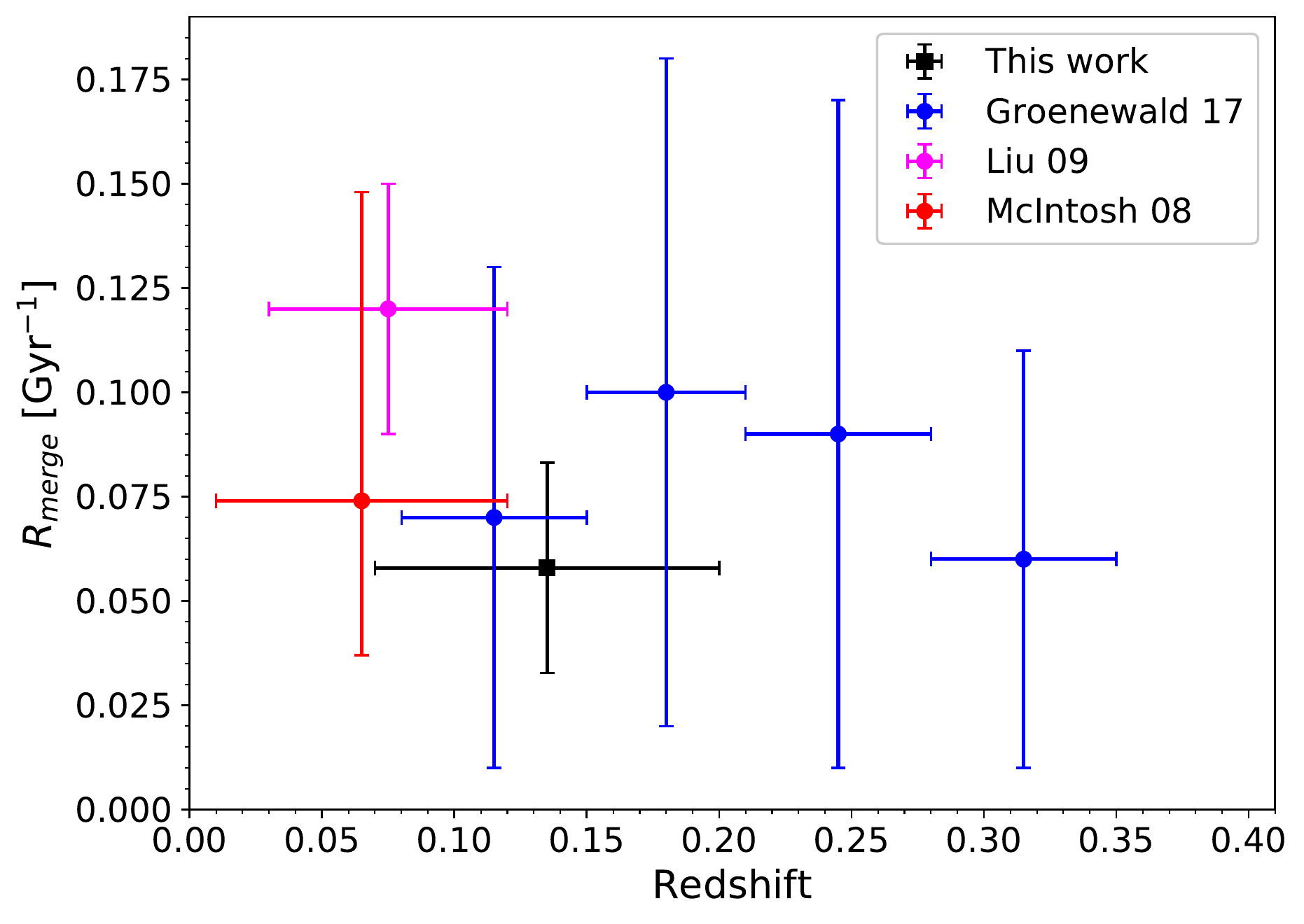}
    \caption{Comparison of the BGG major merger rate. Our result is presented by the black point. Other published merger rates are plotted at the mean redshift of each sample while the horizontal error bars indicate the redshift range of the sample. Our result is consistent with previous studies, that is, BGGs do not experience significant numbers of major mergers per Gyr.}
    \label{fig:Rmerge_compare}
\end{figure}

Figure \ref{fig:Rmerge_compare} shows the comparison of our measured merger rates with the previous studies introduced in Section \ref{BGG_fpair}. While some of these samples have higher stellar mass companions than our sample, we find them all to have consistent merger rates. 

% ---------------NEW SUBSECTION --------------- %
\subsection{Stellar Mass Growth Rate of BGGs} \label{GrowBGG}
The average merger rate and the mean stellar mass of close-companion galaxies are used to estimate the potential stellar mass growth rate of BGGs in our sample. The average stellar mass of the close-companion galaxies in our sample is $8.18\pm3.34\times10^{10}$\,M$_\odot$. The BGGs in our sample therefore increase their stellar mass by $0.47\pm0.30\times10^{10}$\,M$_\odot$\,Gyr$^{-1}$ due to major mergers which is equivalent to a fractional mass increase of $2.19\pm1.52\%$\,Gyr$^{-1}$.

% ---------------NEW SUBSECTION --------------- %
\subsubsection{Observational studies of close-companion galaxies}
\begin{figure}
    \centering
    \includegraphics[width=0.48\textwidth]{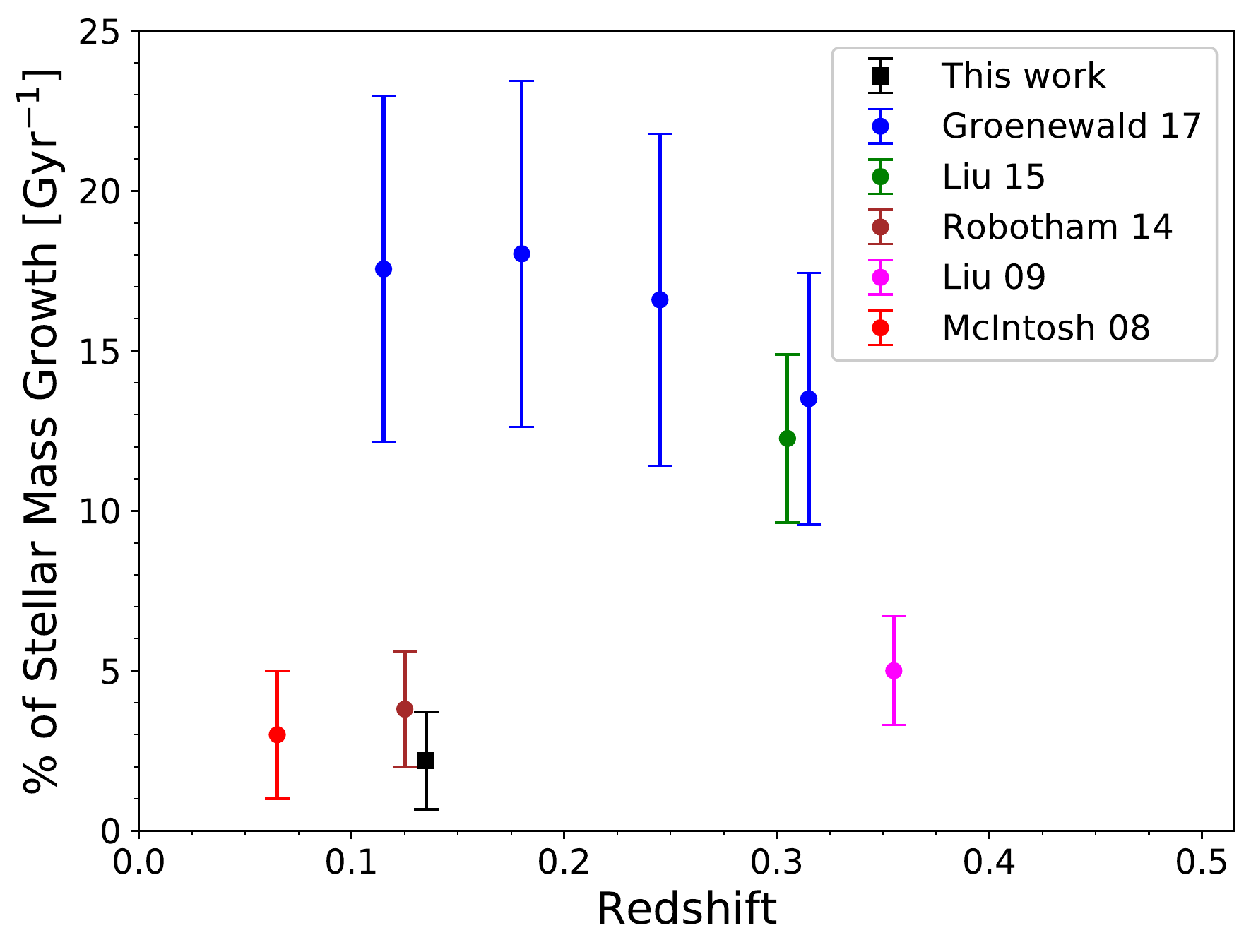}
    \caption{Comparison of the fractional stellar mass growth of BGGs per Gyr due to major mergers with respect to redshift. Our result is presented by the black point. Other published fractional stellar mass growth of BGGs are plotted at the average redshift of each sample. Overall, BGGs in our sample, as well as those from other samples, do not undergo a significant growth rate.}
    \label{fig:growth_compare}
\end{figure}

We compare the values calculated in \cite{McIntosh08,Liu09,Liu15,Groenewald17} with the stellar mass growth we have estimated in Fig. \ref{fig:growth_compare}. It is important to note that only \cite{McIntosh08} have calculated the stellar mass growth of BGGs per Gyr. All other studies mentioned in this comparison estimate the stellar mass growth over a redshift range. In order to obtain a comparison to these results we convert them to a stellar mass growth per Gyr by dividing the stellar mass growth by the lookback-time that corresponds to the redshift range using the same cosmology.

Numerical simulations that investigate the build up of the intracluster light (ICL; e.g. \citealt{Murante07,Conroy07,Puchwein10,Laporte13,Contini14}) due to galaxy mergers predict that $30-80\%$ of a merging close-companion galaxy's stellar mass contributes to the mass of the ICL rather than the BGG. Many observational studies that investigate the stellar mass build up of BGGs take this into account and assume a conservative fraction, $f=0.5$ (e.g. \citealt{Liu09,Liu15,Groenewald17}). This study investigates the potential for stellar mass build up of BGGs, hence we do not assume this mass fraction. Henceforth, we divide out the fractions used in these studies to present a consistent comparison to our result.

\cite{McIntosh08} estimate the rate of stellar mass accretion by major mergers via,
\begin{equation}
    \dot{M}_{\text{BGG}}=\frac{\Sigma M_{*,i}f}{N_{\text{BGG}}}\times\frac{1}{t_{\text{merge}}}
\end{equation}
where $N_{\text{BGG}}$ is the total number of BGGs in the sample; $M_{*,i}$ is the stellar mass of the $i$-th close-companion galaxy galaxy; and $f$ is the fraction of stellar mass of the companion galaxy that is accreted onto the BGG. They find that BGGs in large groups with redshifts $z\leq0.12$ gain up to $2.4^{+1.1}_{-0.6}\times10^{10}$ M$_\odot$ Gyr$^{-1}$ assuming that all of the companion's stellar mass is accreted onto the BGG. The stellar mass growth calculated by \cite{McIntosh08} is equivalent to an average stellar mass growth of $1-5\%$ Gyr$^{-1}$. This is represented in Fig. \ref{fig:growth_compare} by the red point. 

\cite{Liu09} found that major mergers contribute $5.0\pm1.7\%\,$Gyr$^{-1}$ when all of the major companion's stellar mass is accreted onto the BGG. This is represented by the magenta point in Fig. \ref{fig:growth_compare}. In a later study, \cite{Liu15} found that major mergers contribute $70\pm15\%$ to the stellar mass of present day BGGs since $z=0.6$. This corresponds to an average stellar mass growth of $12.2\pm2.6\%$ Gyr$^{-1}$ from $z=0.6$ (green point in Fig. \ref{fig:growth_compare}) which is substantially larger than their earlier estimate of the potential stellar mass growth rate as well as what we calculate here. We will return to this discrepancy later in this section.

\cite{Edwards12} also identify close-companion galaxies of BGGs in order to quantify the rate at which these galaxies grow via mergers. They used deep images from the Canada-France-Hawaii Telescope Legacy Survey and analysed close-companion galaxies within 50 kpc (using photometric redshifts) of their BGGs with luminosity ratios up to $L_{\text{BGG}}/L_{\text{CC}}=20$. The luminosity in major companions is $1.14\pm0.28\times10^{10}\,$L$_{\odot}$, whereas it is almost double when including minor companions, i.e. $2.14\pm0.31\times10^{10}\,$L$_{\odot}$. They find that these close-companion galaxies could increase the stellar mass of a $5\times10^{11}\,$M$_{\odot}$ BGG by up to $\sim10\%$ over the redshift range $0.15\leq z\leq0.39$. This potential stellar mass growth is attributed to both major and minor close-companion galaxies, therefore we do not directly compare it to our result in Fig \ref{fig:growth_compare}.

The stellar mass growth of galaxies from an earlier release of the GAMA Galaxy Group catalogue was also investigated by \cite{Robotham14}. They investigated major mergers of galaxies with stellar masses $10^{8}-10^{12}\,$M$_\odot$ and found that the fraction of mass being added by merging is approximately $2.0-5.6\%$. This is represented in Fig. \ref{fig:growth_compare} by the dark red point.

\cite{Groenewald17} investigated the stellar mass build-up of BGGs between $0.08\leq z\leq0.50$. They found major mergers contribute on average $48\pm17\%$ towards the stellar mass of present day BGGs since $z=0.32$. This corresponds to a stellar mass growth rate of $\sim17\%$ Gyr$^{-1}$. These results are plotted as blue points in Fig. \ref{fig:growth_compare}. They find significantly higher stellar mass growth from major mergers than we do here.

Overall, we find that our results are in good agreement with the stellar mass growth calculated by \cite{McIntosh08,Liu09} and \cite{Robotham14}. However, the growth predicted by \cite{Liu15} and \cite{Groenewald17} is significantly larger. This is not unexpected due to the different methods employed by these studies. These growth rates were calculated with either purely photometric galaxy surveys (e.g. \citealt{McIntosh08,Liu09,Edwards12,Liu15}) or incomplete spectroscopic surveys (e.g. \citealt{Groenewald17}). \cite{McIntosh08} and \cite{Liu09} refined their search for close-companion galaxies by visually inspecting their sample of BGGs for signs of merging. This ensures that the galaxies selected are true close-companion galaxies such that their growth rates are consistent with our robust measurements from a highly-complete spectroscopic survey. The stellar mass growth rates estimated by \cite{Edwards12,Liu15} and \cite{Groenewald17} are larger suggesting that the corrections applied for close-companion galaxies are not conservative enough. 

% ---------------NEW SUBSECTION --------------- %
\subsubsection{Observational studies of change in stellar mass}
Other studies such as \cite{Lidman12} and \cite{Oliva14} estimate the stellar mass growth of BGGs via a direct examination of the change in stellar mass of BGGs from higher to lower redshifts. \cite{Lidman12} find that the stellar mass of BGGs increases by a factor of $1.8\pm0.3$ between $z\sim0.9$ and $z\sim0.2$, approximately equivalent to 5\,Gyr. %This is represented by the dark green point in Fig. \ref{fig:models}.

\cite{Oliva14} also directly examined the change in stellar mass of BGGs at low and high redshifts, $0.09\leq z\leq0.27$ using a large sample of 883 galaxies from an earlier release of the GAMA survey. They found an average stellar mass growth from $z=0.27$ to $z=0.09$ of $-7\pm9\%$. This suggests that BGGs do not grow significantly over this redshift range. Our result agrees that BGGs do not grow in stellar mass significantly at low redshifts. It is important to note that while these results are drawn from the same survey, the stellar mass growth of the BGGs are estimated with independent methods.

% ---------------NEW SUBSECTION --------------- %
\subsubsection{Stellar mass growth in models}
Semi-analytical models (e.g. \citealt{DeLucia07,Laporte13,Contini14}) used the Millennium Simulation \citep{Springel05} to study the formation and evolution of BGGs. The results of these models are largely similar. Both \cite{DeLucia07} and \cite{Laporte13} find that BGGs grow by a factor of $\sim1.8$ at low redshifts ($z<0.5$), which is similar to that found observationally by \cite{Lidman12}. This is equivalent to a growth in stellar mass of $\sim15\%$ per Gyr, which is significantly larger than what we find observationally. However, further evolution of the model in \cite{DeLucia07} explored in \cite{Contini14}, which accounts for the parallel growth of the intracluster light from massive merging galaxies, find that the intracluster light fraction in groups and clusters ranges between 20--40\%. Central galaxies in this updated model do not grow significantly at low redshifts, having a mass of approximately 97\% of their total mass at $z\sim0.2$. This agrees with the results obtained here. 

% ---------------NEW SECTION --------------- %
\section{Conclusion} \label{sec:conclusion}
In this paper we have examined the potential for stellar mass build-up of BGGs in the local Universe due to major mergers. We have analysed a large volume-limited sample of 550 groups with spectroscopic redshifts between $0.07\leq z<0.2$ from the Galaxy And Mass Assembly (GAMA) survey \citep{Liske15}. This data set is highly spectroscopically complete with a survey magnitude limit of $m_r>19.8$ mag which allows us to robustly analyse the impact of merging companion galaxies.

We selected a volume-limited sample of groups identified within the GAMA Galaxy Group Catalogue \citep{Robotham11}. The BGGs studied here lie at the centre of these groups and possess stellar masses between $10^{11.0}$\,M$_\odot$ and $10^{11.6}$\,M$_\odot$ where the stellar masses are sourced from the GAMA Stellar Mass catalogue \citep{Taylor11}.

We identified close-companion galaxies within a projected radius of 30\,kpc of the central BGG with relative velocities $\Delta\leq300$\,km\,s$^{-1}$ as a proxy to estimate the potential for stellar mass growth of BGGs within our sample. We investigated potential major merger candidates with mass ratios $M_{\text{BGG}}/M_{\text{CC}}\leq4$.

Within our sample of 550 groups we identified 17 close-companion galaxies. This resulted in a total pair fraction of $0.03\pm0.01$ with no significant evolution over the redshift range studied here. We also investigated the dependence on halo mass and, while we found no statistically significant dependence, we do observe a systematic decrease in the BGG pair fraction with increasing halo mass.

The close-companion galaxies in our sample will merge with their respective BGGs in $T_{\text{merge}}=0.53\pm0.19\,$Gyr, estimated using the merging timescale derived in \cite{Kitzbichler08}. This combined with the pair fraction resulted in an average merger rate of $0.06\pm0.03$\,Gyr$^{-1}$ for BGGs in our sample. This results in a potential average stellar mass growth of $2.19\pm1.52\%$\,Gyr$^{-1}$ over $0.07<z<0.2$ due to major mergers. This is in good agreement with other observational studies that investigate the stellar mass build up of BGGs via mergers, however, it is lower than that predicted by semi-analytical models. 

In a future analysis we will use recent cosmological simulations such as \textsc{Illustris}, EAGLE, and \textsc{Simba} (\citealt{Vogelsberger14,Schaye15,Dave19} respectively) to directly compare with these observational results to further validate the use of close-companion galaxies as a proxy for BGG stellar mass growth.

\acknowledgments

The authors thank the anonymous referee for their comments which improved the paper.

GAMA is a joint European-Australasian project based around a spectroscopic campaign using the Anglo-Australian Telescope. The GAMA input catalogue is based on data taken from the Sloan Digital Sky Survey and the UKIRT Infrared Deep Sky Survey. Complementary imaging of the GAMA regions is being obtained by a number of independent survey programmes including GALEX MIS, VST KiDS, VISTA VIKING, WISE, Herschel-ATLAS, GMRT and ASKAP providing UV to radio coverage. GAMA is funded by the STFC (UK), the ARC (Australia), the AAO, and the participating institutions. The GAMA website is http://www.gama-survey.org/ .

The authors acknowledge the Traditional Custodians and the Elders past and present of the land in which the Anglo-Australian Telescope stands, the Gamilaraay people, and the Bedegal people who are the Traditional Custodians of the land, of which much of this research was conducted, at UNSW Sydney. 

The authors would like to thank C. Conselice for his invaluable comments during the preparation of this paper. 

SB acknowledges funding support from the Australian Research Council through a Future Fellowship (FT140101166). AR acknowledges funding support from the Australian Research Council through a Future Fellowship (FT200100375). 

% \bibliography{paper}{}
\bibliographystyle{aasjournal}

\end{document}